\documentclass[aps,prb,onecolumn,superscriptaddress,showpacs]{revtex4}
\usepackage{graphicx}
\usepackage{latexsym}
\usepackage{amssymb}
\usepackage{amsmath}
\usepackage{amsfonts}
\usepackage{bm}
\usepackage{multirow}
\usepackage{color}
\usepackage{comment}
\newcommand{\ii}{\mathrm{i}}

\newcommand{\bx}{\mathbf{x}}

\newcommand{\he}{\hat{e}}
\newcommand{\hve}{\hat{\vec{e}}}

\newcommand{\va}{\vec{a}}
\newcommand{\ha}{\hat{a}}
\newcommand{\hva}{\hat{\vec{a}}}

\newcommand{\cA}{\mathcal{A}}
\newcommand{\cC}{\mathcal{C}}

\newcommand{\hB}{\hat{\mathcal{B}}}

\newcommand{\hn}{\hat{n}}
\newcommand{\htheta}{\hat{\theta}}
\newcommand{\hN}{\hat{N}}
\newcommand{\hphi}{\hat{\phi}}
\newcommand{\vn}{\vec{\nabla}}

\newcommand{\tO}{\tilde{O}}
\newcommand{\tZ}{\tilde{Z}}
\newcommand{\tU}{\tilde{U}}
\newcommand{\ti}{\tilde{i}}
\newcommand{\tj}{\tilde{j}}
\newcommand{\tx}{\tilde{x}}
\newcommand{\ty}{\tilde{y}}
\newcommand{\tz}{\tilde{z}}
\newcommand{\tl}{\tilde{l}}
\newcommand{\hmu}{\hat{\mu}}
\newcommand{\hnu}{\hat{\nu}}
\newcommand{\hrho}{\hat{\rho}}
\newcommand{\hx}{\hat{x}}
\newcommand{\hy}{\hat{y}}
\newcommand{\hz}{\hat{z}}
\newcommand{\btx}{\tilde{\mathbf{x}}}

\newcommand{\U}{\mathrm{U}}

\newcommand{\beq}{\begin{equation}}
\newcommand{\eeq}{\end{equation}}
\newcommand{\beqn}{\begin{eqnarray}}
\newcommand{\eeqn}{\end{eqnarray}}

\def\U{{\rm U}}

\begin{document}
\title{Categorical Symmetries at Criticality}

\author{Xiao-Chuan Wu}
\affiliation{Department of Physics, University of California,
Santa Barbara, CA 93106}

\author{Wenjie Ji}
\affiliation{Department of Physics, University of California,
Santa Barbara, CA 93106}

\author{Cenke Xu}
\affiliation{Department of Physics, University of California,
Santa Barbara, CA 93106}

\begin{abstract}

We study the concept of ``categorical symmetry" introduced
recently, which in the most basic sense refers to a pair of dual
symmetries, such as the Ising symmetries of the $1d$ quantum Ising
model and its self-dual counterpart. In this manuscript we study
discrete categorical symmetry at higher dimensional critical
points and gapless phases. At these selected gapless states of
matter, we can evaluate the behavior of categorical symmetries
analytically. We analyze the categorical symmetry at the following
examples of criticality: ({\it i.}) $(2+1)d$ Lifshit critical
point of a quantum Ising system; ({\it ii.}) $(3+1)d$ photon phase
as an intermediate gapless phase between the topological order and
the confined phase of $3d$ $Z_2$ quantum gauge theory; ({\it
iii.}) $2d$ and $3d$ examples of systems with both categorical
symmetries (either 0-form or 1-form categorical symmetries) and
subsystem symmetries. We demonstrate that at some of these gapless
states of matter the categorical symmetries have very different
behavior from the nearby gapped phases.

\end{abstract}

\date{\today}

\maketitle

\section{Basics of categorical symmetry}

Categorical symmetry is a new concept introduced in
Ref.~\onlinecite{jicsym}, which expanded the conventional notion
of symmetries in physics, and how one should think about them. The
basic examples of categorical symmetry correspond to a pair of
dual symmetries, whose local symmetry charges in general do not
commute with each other. The simplest example of such, are the
$Z_2$ and $\tZ_2$ dual symmetry of the $1d$ quantum Ising model:
\beqn H = \sum_j - K \sigma^3_j \sigma^3_{j+1} - h \sigma^1_j \ \
\leftrightarrow \ \ H_d = \sum_{\tj} - K \tau^1_{\tj} - h
\tau^3_{\tj}\tau^3_{\tj + 1}. \eeqn This model has a well-known
self-duality point $K = h$; $\sigma^3_j$ and $\tau^3_{\tj}$ are
order parameters of the original $Z_2$ and the dual $\tZ_2$
symmetry. Let us label the entire categorical symmetries of the
$1d$ quantum Ising model as $Z_2 \star \tilde{Z}_2$.

For the convenience of generalizing to higher dimensional systems
with higher form symmetries and more exotic subsystem symmetries
that we will discuss in this manuscript, we will introduce the
concept ``Order Diagnosis Operator" (ODO) for each symmetry. The
expectation value of the ODO diagnoses the behavior of its
corresponding symmetry. An ODO should commute with all the {\it
conserved global} symmetry charges (which implies that the
expectation value of the ODO is in general nonzero~\footnote{The
expectation value of ODOs should not be viewed as an analogue of
order parameter, they should be viewed as analogue of correlation
of order parameters. The ODOs were studied as the ``patch symmetry
operators" of the categorical symmetry in
Ref.~\onlinecite{jicsym}.}), but creates {\it local} charges of
the corresponding symmetry. For the $Z_2$ and $\tZ_2$ symmetries
of the $1d$ quantum Ising model, the ODOs are respectively \beqn
O_{i,j} = \sigma^3_i \sigma^3_j, \ \ \ \tO_{\ti, \tj} =
\tau^3_{\ti} \tau^3_{\tj} = \prod_{i<k<j} \sigma^1_k. \eeqn
$O_{i,j}$ creates a pair of $Z_2$ charges at sites $i$ and $j$
(but it preserves/commutes with the global $Z_2$ charge $\prod_j
\sigma^1_j$), while $\tO_{\ti, \tj}$ creates a pair of domain
walls of $\sigma^3$ at $\ti$ and $\tj$, which are local charges of
the $\tZ_2$ symmetry.

When $K > h$, there is a long range correlation of $\sigma^3$,
short range correlation of $\tau^3$ (long range expectation value
of ODO $O_{i,j}$, and short range expectation value of $\tO_{\ti,
\tj}$); hence this is a phase that spontaneously breaks $Z_2$, but
preserves $\tZ_2$. When $K < h$, there is a long range correlation
of $\tau^3$, but short range correlation of $\sigma^3$ (long range
expectation value of $\tO_{\ti, \tj}$, short range expectation
value of $O_{i,j}$); hence this is a phase that spontaneously
breaks $\tZ_2$, but preserves $Z_2$. Whether a symmetry is
preserved or spontaneously broken, can be defined by the behavior
of its ODO. When $K = h$, both order parameters have power-law
correlation, hence this is a criticality which preserves both
symmetries.

In what sense is $\tZ_2$ a symmetry, and in what sense is there a
spontaneous symmetry breaking (SSB) of $\tZ_2$? In the $1d$
quantum Ising model, without changing the physical Ising Hilbert
space, the SSB phase of the $\tZ_2$ symmetry does not lead to
ground state degeneracy (GSD), after all it is just a quantum
disordered phase of the Ising model. However, with some global
constraint on the physical Hilbert space, or when we view the $1d$
system as the boundary of a $2d$ topological order~\cite{jicsym},
neither phase ($K > h$ or $K < h$) has GSD. {\it Hence we no
longer view GSD as a criterion for SSB. The SSB should be defined
solely by the behavior of $\langle O \rangle$ and $\langle \tO
\rangle$.}

%The behavior associated with $\tZ_2$ is also clearly distinct with
%systems without any symmetry, because without any symmetry, the
%correlation function of any ordinary operators should always be
%long ranged (like the Ising order parameter when the Ising
%symmetry is explicitly broken); while $\tO_{\ti,\tj}$ can indeed
%have long range, short range, and power-law behaviors in different
%phases.

In higher dimensions, the possible categorical symmetries are much
richer. In the $2d$ quantum Ising model, there is a $Z_2 \star
\tZ^{(1)}_2$ symmetry. Here $\tZ_2^{(1)}$ is a 1-form symmetry as
a generalization of ordinary symmetries introduced in recent years
(see for instance
Ref.~\onlinecite{formsym1,formsym2,formsym3,formsym4,formsym5,formsym6,formsym7,formsym8,Cordova2019}):
\beqn H &=& \sum_{<\bx,\bx'>} - K \sigma^3_{\bx}\sigma^3_{\bx'} -
\sum_\bx h \sigma^1_\bx \ \ \leftrightarrow \crcr H_d &=&
\sum_{\btx, \hat{\mu} } - K \tau^1_{\btx, \hat{\mu}} - \sum_{\btx}
h \tau^3_{\btx, \hx} \tau^3_{\btx, \hy} \tau^3_{\btx + \hat{x},
\hy} \tau^3_{\btx + \hat{y}, \hx}. \label{2dz2gauge}\eeqn The
lattice site $\bx$ and dual lattice site $\btx$ are illustrated in
Fig.~\ref{clattice}. The subscripts $(\btx, \hx)$ and $(\btx,
\hy)$ label the links of the dual lattice. The ODO of the $Z_2$
symmetry is still $O_{\bx,\bx'} = \sigma^3_\bx \sigma^3_{\bx'}$;
while the ODO of $\tZ_2^{(1)}$ symmetry is \beqn
\tO^{(1)}_\mathcal{\mathcal{C}} = \prod_{\tl \in \mathcal{C}}
\tau^3_{\tl} = \prod_{\bx \in \mathcal{A}, \
\partial \mathcal{A} = \mathcal{C}} \sigma^1_\bx. \eeqn Here $\tl$
also labels a link in the dual lattice, which belongs to the
contractible loop $\cC$. $\tO^{(1)}_\mathcal{C}$ creates an Ising
domain wall of $\sigma^3$, the one dimensional domain wall carries
the dual $\tZ_2^{(1)}$ 1-form symmetry charge. Here $\mathcal{A}$
is a finite $2d$ patch on the dual lattice, $\mathcal{C}$ is the
boundary of $\mathcal{A}$, which is a contractible loop. Again,
the ODO $\tO^{(1)}_\mathcal{C}$ commutes with all the conserved
1-form symmetry charges, which is defined as a product of $\tau^1$
along any closed $1d$ loop $\cC'$. Notice that $\cC'$ always
intersects with the contractible $\cC$ for even times, hence the
ODO $\tO^{(1)}_\mathcal{C}$ commutes with the conserved 1-form
symmetry charges $\prod_{\tl \in \cC'} \tau^1_{\tl}$.

There are again two phases with $K/h$ greater or smaller than a
critical value. These two phases have the following known
behaviors of the ODOs~\cite{susskind}, which can be computed
through a reliable perturbation theory due to the gap in the
spectrum of both phases: \beqn && K/h \gg 1, \ \ \langle O_{\bx,
\bx'} \rangle \sim \mathrm{Const}, \ \ \ \langle
\tO^{(1)}_{\mathcal{C}} \rangle \sim e^{- \alpha_1 \log(K/h)
\mathcal{A}}. \cr\cr && K/h \ll 1, \ \ \langle O_{\bx, \bx'}
\rangle \sim e^{- |\bx - \bx'| / \xi}, \ \ \ \langle
\tO^{(1)}_{\mathcal{C}} \rangle \sim e^{- \alpha_2 (K/h)^2
\mathcal{C}}. \eeqn $\alpha_i$ are order 1 numbers. Hence in the
phase $K \gg h$, the $\tZ_2^{(1)}$ symmetry $\tO^{(1)}_{\cC}$
decays with an area law; while in the phase $K \ll h$, the domain
walls proliferate/condense, and $\tO^{(1)}_{\cC}$ has a perimeter
law. Again, in the phase $h \gg K$, even though the domain walls
proliferate/condense, there is no GSD. This is in stark contrast
with ordinary 1-form symmetry SSB state, which would lead to
topological degeneracy. Hence here we should view the behavior of
$\langle \tO^{(1)}_{\cC} \rangle$ as a criterion of SSB of
$\tZ_2^{(1)}$, rather than the GSD.

At the $(2+1)d$ Ising critical point, the $Z_2$ order parameter
has a power-law correlation (the expectation value of
$O_{\bx,\bx'}$ falls off as a power-law), hence the $Z_2$ symmetry
is not broken. Intuitively, since $O_{\bx,\bx'}$ has a power-law
correlation, the expectation value of the dual ODO
$\tO^{(1)}_\mathcal{C}$ should be stronger than the area law deep
in the $K \gg h$ phase, but weaker than the perimeter law deep in
the $K \ll h$ phase. But the exact behavior of
$\tO^{(1)}_\mathcal{C}$ is difficult to compute analytically at
the $3D$ Ising critical point, and in other lattice models that
will be discussed in the following sections. The main goal of this
manuscript is to find critical points (or fine-tuned critical
points) where the ODOs of the categorical symmetries can be
evaluated analytically. The strategy we will generally take is
that, we embed the target lattice model into a larger ``parent"
system where the ODOs of the original system have a clear
representation. Then we tune the parent system to a multi-critical
point, or even a gapless phase, where we can use tools in the
continuum limit to compute ODOs defined in both sides of the
duality. Since many of the states we discuss in this manuscript do
not have Lorentz invariance, we will focus on expectation value of
time-independent operators at static states.

\section{Ising Categorical symmetries at criticality}

\subsection{$2d$ Lifshitz Point}

\begin{figure}
\includegraphics[width=0.5\textwidth]{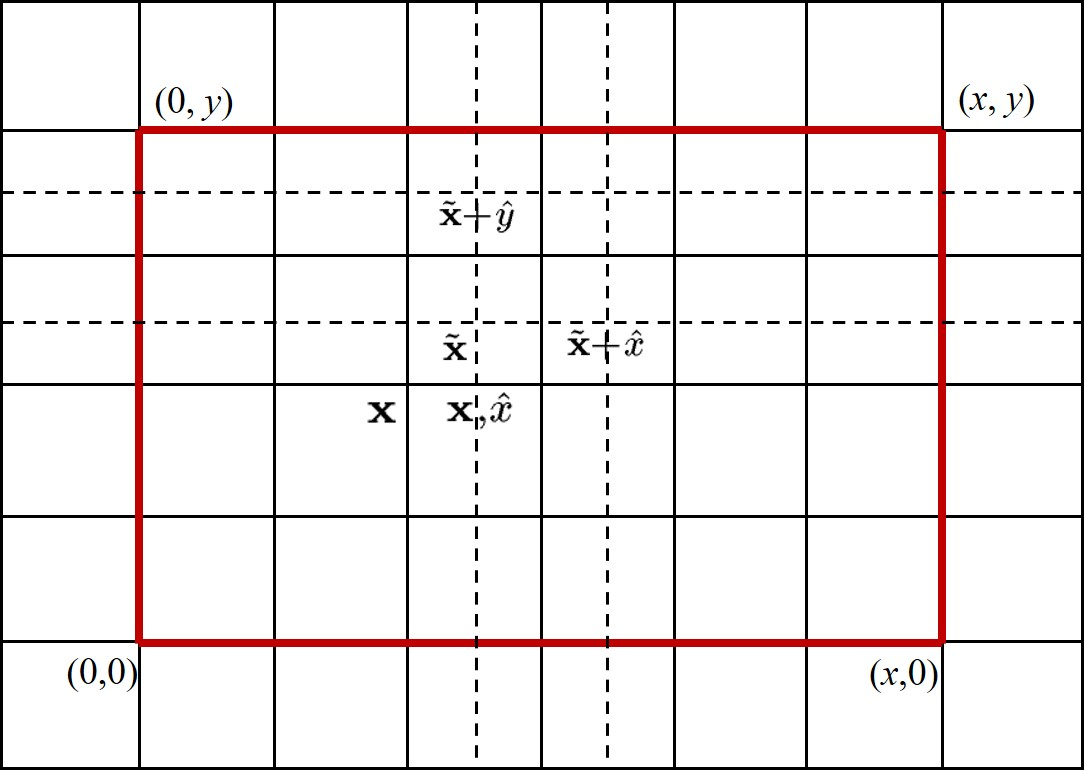}
\caption{The $2d$ square lattice, and its dual lattice. The
lattice site is labelled as $\bx$, and the dual lattice site (the
plaquette of the original lattice) is labelled as $\btx$. The
links of the lattice are labelled as $(\bx, \hmu)$, while the
links of the dual lattice are labelled as $(\btx, \hmu)$. }
\label{clattice}
\end{figure}

We can embed the target $2d$ quantum Ising model into a parent
system described by a $\U(1)$ quantum ``rotor": \beqn H =
\sum_{\bx,\mu} - t \cos\left(\nabla_\mu \htheta(\bx)\right) +
\sum_\bx \frac{U}{2}\hn(\bx)^2 - g \cos\left(2\htheta(\bx)\right).
\label{u12d}\eeqn $\htheta(\bx)$ and $\hn(\bx)$ are a pair of
conjugate variables, i.e. $[\hn(\bx), \htheta(\bx')] =
\ii\delta_{\bx,\bx'}$. $\hn(\bx)$ takes discrete integer
eigenvalues, while $\htheta(\bx)$ is periodically defined:
$\htheta(\bx) = \htheta(\bx) + 2\pi$. The last $g$ term in
Eq.~\ref{u12d} breaks the $\U(1)$ symmetry down to $Z_2$. The
operators $\sigma^3_\bx$ and $\sigma^1_\bx$ of the Ising model
correspond to the operators in the parent $\U(1)$ theory: \beqn
\sigma^3_\bx = e^{\ii \htheta(\bx) }, \ \ \ \sigma^1_\bx = e^{\ii
\pi \hn(\bx)}. \eeqn If the $g$ term is ignored, the $\U(1)$ model
is dual to a lattice QED: \beqn && H_d = \sum_{\btx} - t
\cos\left(\hve(\btx)\right) + \sum_{\btx} \frac{U}{2}\left(\vn
\times \hva(\btx) \right)^2 \cr\cr && \hve(\btx) = \hat{z} \times
\vn \htheta(\bx), \ \ \ \vn \times \hva(\btx) = \hn(\bx). \eeqn
The electric field $\he_\mu$ and gauge vector potential $\ha_\mu$
were defined on the links $(\btx, \hat{x})$, $(\btx, \hat{y})$ of
the dual lattice, but we can also equivalently define $\hve(\btx)
= (\he_x(\btx), \he_y(\btx)) = (\he_{\btx, \hx},\he_{\btx, \hy})$,
$\hva(\btx) = (\ha_x(\btx), \ha_y(\btx)) = (\ha_{\btx,
\hx},\ha_{\btx, \hy})$. In the parent $\U(1)$ system, the $Z_2$
and $\tilde{Z}_2^{(1)}$ ODO are \beqn && O_{\bx,\bx'} = e^{\ii
\htheta(\bx) } e^{- \ii \htheta(\bx')} , \cr \cr &&
\tO^{(1)}_\mathcal{C} = \prod_{\mathcal{A}, \
\partial \mathcal{A} = \mathcal{C}} \sigma^1_{\bx} = \exp\left( i \pi
\sum_{\bx \in \mathcal{A}} \hn(\bx ) \right) = \exp \left( i\pi
\oint_\mathcal{\mathcal{C}} \hva \cdot d\vec{l} \right). \eeqn

In model Eq.~\ref{u12d}, there is a critical point at critical
value $(U/t)_c$. Without the $g$ term, the transition in
Eq.~\ref{u12d} is a $3D$ XY transition between the superfluid
phase with small $U/t$ and a boson Mott insulator phase at large
$U/t$. While with the $g$ term, it is expected that the $3D$ XY
critical point will flow to the $3D$ Ising fixed point, because
$g$ is obviously relevant at the $3D$ XY fixed point. However, one
can fine-tune the critical point to reach a Lifshitz point
described by the following field theory Hamiltonian and action in
the continuum limit \beqn && H = \int d^2x \ \frac{U}{2}
\hn(\bx)^2 + \frac{\rho}{2} \left(\nabla^2 \htheta(\bx)\right)^2,
\cr\cr && \mathcal{S} = \int d^2x d\tau \ \frac{1}{2U}
(\partial_\tau \theta)^2 + \frac{\rho}{2} (\nabla^2 \theta)^2.
\label{lif}\eeqn It is known that the $g$ operator can be
irrelevant at the $(2+1)d$ Lifshitz Gaussian fixed point for
certain range of $U$ and $\rho$, more precisely for large enough
$U/\rho$~\cite{henley,fradkin2004}. The irrelevance of $g$
guarantees that the continuum limit field theory description in
terms of $\theta$ is applicable at this Lifshitz fixed point. One
can also compute the expectation value of $O$, which is the
equal-time correlation function between $\sigma^3$: \beqn \langle
O_{0,\bx} \rangle = \langle e^{\ii \htheta(0)}e^{- \ii
\htheta(\bx)} \rangle \sim \frac{1}{|\bx|^{2\Delta_\theta}}, \ \ \
\Delta_\theta \sim \sqrt{\frac{U}{\rho}}. \eeqn Hence at the
Lifshitz point, the $Z_2$ symmetry is preserved.

The situation is rather different for the $Z_2^{(1)}$ ODO
$\tilde{O}_\mathcal{C}$. The dual Hamiltonian and action of the
Lifshitz theory Eq.~\ref{lif} is \beqn && H_d = \int d^2 \tx \
\frac{U}{2} \left( \vec{\nabla} \times \hva \right)^2 +
\frac{\rho}{2} \left( (\nabla_x \he_y)^2 +(\nabla_y \he_x)^2
\right), \cr\cr && \mathcal{S}_d = \int d^2 \tx d\tau \
\frac{1}{2\rho} \left( \ha_x \frac{\partial_\tau^2}{ \partial_y^2
} \ha_x + \ha_y \frac{\partial_\tau^2}{ \partial_x^2 } \ha_y
\right) + \frac{U}{2} (\vec{\nabla}\times \hva)^2. \eeqn This is
the same Hamiltonian and action describing the $2d$ quantum dimer
model at the Rohksar-Kivelson point~\cite{rk,dimergauge}. The
correlation function of $\va_{\vec{q},\omega}$ is \beqn \langle
\ha_{\mu} (-\omega, - \vec{q}) \ha_{\nu} (\omega, \vec{q}) \rangle
\sim \frac{\rho(q^2 \delta_{\mu\nu} - q_\mu q_\nu)}{\omega^2 +
\rho U q^4}, \ \ \ \langle \ha_\mu(0, 0) \ha_\nu (0, \mathbf{x})
\rangle \sim \sqrt{\frac{\rho}{U}}\frac{1}{|\mathbf{x}|^2}. \eeqn

The expectation of $\tO^{(1)}_\mathcal{C}$ can be evaluated using
the Gaussian theory of the gauge field: \beqn \langle \exp(\ii \pi
\oint_\mathcal{C} \hva \cdot d\vec{l}) \rangle \sim \exp\left( -
\frac{\pi^2}{2} \oint_\mathcal{C} \oint_\mathcal{C} \langle
\ha_\mu (\mathbf{x}) \ha_\nu(\mathbf{x'}) \rangle dx^\mu dx'^{\nu}
\right). \eeqn Power-counting suggests that this is still a
perimeter law: the $1/|\bx|^2$ decay of the correlation function
of the gauge fields do not lead to extra divergence with large
loop size, the expectation value of $\tO^{(1)}_\cC$ is dominated
by small distance correlation of the gauge field. Since in the
gapped phase $h \gg K$ (Eq.~\ref{2dz2gauge}) where the domain
walls clearly proliferates, $\tO^{(1)}_\cC$ follows a perimeter
law, we will use the perimeter law of $\tO^{(1)}_\cC$ as a
criterion of SSB of $\tZ_2^{(1)}$. Then this Lifshitz point still
spontaneously breaks the $\tilde{Z}_2^{(1)}$ symmetry, while
preserving the $Z_2$ symmetry. One can also see that when the
expectation value of $O_{\bx,\bx'}$ is stronger (smaller
$\Delta_\theta$ at smaller $U/\rho$), the expectation value of
$\tO^{(1)}_\mathcal{C}$ becomes weaker (larger $\rho/U$). The
results of this section are summarized in the table below.

\begin{center}
\begin{tabular}{ |c|c|c|c|}
 \hline  $2d$ Quantum Ising theory  & $K \gg h$ in Eq.~\ref{2dz2gauge} & $K \ll h$ in Eq.~\ref{2dz2gauge}  & Fine-tuned Lifshitz Point
 \\
 \hline
 $O_{\bx,\bx'}$  & Long range & Short Range & Power law
 \\
 \hline
 $\tO^{(1)}_\cC$ & Area law & Perimeter law & Perimeter law
 \\
 \hline
\end{tabular}
\label{tablelifshitz}
\end{center}

\subsection{3d $Z_2$ Quantum Gauge Theory}

It was well-known that the $3d$ lattice $Z_2$ gauge theory has a
self-dual structure~\cite{susskind,Kogut1979,savit}: \beqn H &=&
\sum_{\bx, \hmu, \hnu} - K \sigma^3_{\bx, \hmu} \sigma^3_{\bx,
\hnu} \sigma^3_{\bx + \hmu, \hnu} \sigma^3_{\bx + \hnu, \hmu} - h
\sigma^1_{\bx, \hmu} \cr\cr \leftrightarrow \ \ H_d &=&
\sum_{\btx, \hmu, \hnu} - K \tau^1_{\btx, \hmu} - h \tau^3_{\btx,
\hmu} \tau^3_{\btx, \hnu} \tau^3_{\btx + \hmu, \hnu} \tau^3_{\btx
+ \hnu, \hmu}. \label{z23d1}\eeqn This system has a $Z^{(1)}_2
\star \tZ_2^{(1)}$ categorical symmetry. The ODOs for $Z^{(1)}_2$
and $\tZ_2^{(1)}$ are \beqn O^{(1)}_\cC = \prod_{l \in \cC}
\sigma^3_{l}, \ \ \ \tO^{(1)}_\cC = \prod_{\tl \in \cC}
\tau^3_{\tl}. \label{ODO3d}\eeqn The $O^{(1)}_\cC$ and
$\tO^{(1)}_{\cC}$ are products of the $K$ and $h$ terms of
Eq.~\ref{z23d1} within $2d$ patch $\mathcal{A}$ with $\partial
\mathcal{A} = \cC$.

There are two phases of this model: for $K \gg h$, $\langle
O^{(1)}_\cC \rangle$ decays with a perimeter law, while $\langle
\tO^{(1)}_\cC \rangle $ decays with an area law; this is a phase
with SSB of $Z_2^{(1)}$ but preserves $\tZ_2^{(1)}$.

In the opposite limit $h \gg K$, $\langle O^{(1)}_\cC \rangle$
decays with an area law, while $\langle \tO^{(1)}_\cC \rangle $
decays with an perimeter law; this is the phase with SSB of
$\tZ_2^{(1)}$ but preserves $Z_2^{(1)}$.

Unfortunately, model Eq.~\ref{z23d1} does not have a second order
transition between the two phases, hence there is no critical
point in model Eq.~\ref{z23d1} where $Z_2^{(1)}$ and $\tZ_2^{(1)}$
are on equal footing. But we can embed the $Z_2$ gauge theory
Eq.~\ref{z23d1} into a QED model with $U(1)^{(1)} \star
\tU(1)^{(1)}$ symmetries, and this QED model has a gapless photon
phase. In this gapless photon phase, both $O^{(1)}_\cC$ and
$O^{(1)}_{\cC'}$ in Eq.~\ref{ODO3d} can be computed using the
Gaussian fixed point theory of the $U(1)$ gauge field, and its
self-dual $\tU(1)$ gauge field. The Gaussian theory of the $U(1)$
and $\tU(1)$ gauge bosons indicates that both $O_{\cC}$ and
$\tO_{\cC}$ follow a perimeter law. Since in the gapped phases of
Eq.~\ref{z23d1} $O_{\cC}$ and $\tO_{\cC}$ at most have a perimeter
law, we view the gapless photon phase of the $U(1)$ gauge field as
a phase which spontaneously breaks both $Z_2^{(1)}$ and
$\tZ_2^{(1)}$ symmetries. This gapless QED would still have
$Z_2^{(1)} \star \tZ_2^{(1)}$ as the UV symmetry, while the
$U(1)^{(1)} \star \tU(1)^{(1)}$ symmetry are IR emergent
symmetries. The IR emergent symmetries are spontaneously broken,
which still leads to gapless photons as their Goldstone
modes~\footnote{Spontaneous breaking of emergent higher form
symmetries in the infrared would still lead to gapless Goldstone
modes, this is very different from the scenario of ordinary 0-form
symmetries.}.

One can also fine-tune the QED to a Lifshitz point with
non-Lorentz invariant dispersions of the $U(1)$ gauge bosons.
However, we have checked and verified that, at various Lifshitz
points (meaning fine-tuned states with different non-Lorentz
invariant dispersion), at least one of the $Z_2^{(1)}$ and
$\tZ^{(1)}_2$ symmetries is spontaneously broken, i.e. one of
$O_{\cC}$ and $\tO_{\cC}$ must have a perimeter law.

\section{Examples of subsystem categorical symmetries}

\subsection{2d Example}

Let us consider a special $2d$ lattice $Z_2$ quantum gauge theory,
which can be constructed in Josephson arrays of superconductor and
ferromagnet deposited on top of a quantum spin Hall
insulator~\cite{xufu}: \beqn H = \sum_{\bx} -K \sigma^3_{\bx,
\hx}\sigma^3_{\bx, \hy}\sigma^3_{\bx + \hx, \hy} \sigma^3_{\bx +
\hy, \hx} - J \sigma^1_{\bx, \hx}\sigma^1_{\bx + \hat{x}, \hx}- J
\sigma^1_{\bx, \hy}\sigma^1_{\bx + \hat{y}, \hy}. \label{z22d1}
\eeqn The last two terms of this model are actually identical, due
to the $Z_2$ Gauss law gauge constraint $\sigma^1_{\bx - \hx, \hx}
\sigma^1_{\bx, \hx} \sigma^1_{\bx - \hy, \hy} \sigma^1_{\bx, \hy}
= 1$, which we will impose strictly on the Hilbert space of the
system.

This model has an ordinary $Z^{(1)}_2$ 1-form symmetry, and extra
$Z_{2}^{(\mathrm{sub})}$ subsystem symmetries. The subsystem
symmetry grants the system a series of conserved quantities: \beqn
\Sigma_{\hx, y} = \prod_{y = \mathrm{Const}} \sigma^3_{\bx, \hx},
\ \ \ \ \Sigma_{\hy, x} = \prod_{x = \mathrm{Const}}
\sigma^3_{\bx, \hy}. \eeqn $x$ and $y$ are the two coordinates of
$\bx$. The subsystem symmetries of Eq.~\ref{z22d1} guarantee that
$\Sigma_{\hx,y}$ and $\Sigma_{\hy,x}$ are conserved for arbitrary
$x$ and $y$. The ODO for $Z_2^{(1)}$, and its expectation value in
the topological ordered phase $K \gg J$ is \beqn
O^{(1)}_{\mathcal{C}} = \prod_{l \in \cC} \sigma^3_l, \ \ \ \
\langle O^{(1)}_{\mathcal{C}} \rangle \sim e^{- \alpha_3 (J/K)^2
N_\cC}. \label{corner} \eeqn The $O^{(1)}_{\mathcal{C}}$ commutes
with conserved quantities $\Sigma_{\hx,y}$ and $\Sigma_{\hy,x}$,
hence it meets the criterion of ODO we introduced in the first
section. Due to the conservation of the extra quantities
$\Sigma_{\hx,y}$ and $\Sigma_{\hy,x}$, the ODO has a generic
``corner law" instead of perimeter law, where $N_\cC$ is the
number of corners of loop $\cC$. For example, in
Fig.~\ref{clattice}, the rectangular loop $\cC$ has four corners,
And $O^{(1)}_\cC$ is a product of finite segments of
$\Sigma_{\hx,y}$ and $\Sigma_{\hy,x}$. The expectation value of
the rectangular $O^{(1)}_{\mathcal{C}}$ does not decay with the
length of $\cC$. Because $\Sigma_{\hx,y}$ and $\Sigma_{\hy,x}$ are
conserved when the product is along an infinitely straight line,
then for a generic $\cC$, if we compute the expectation value of
$O^{(1)}_{\cC}$ through a perturbation of $J/K$ like
Ref.~\onlinecite{susskind}, the value can only decay when $\cC$
``takes a turn".

In the other limit of the model, $K \ll J$, the ODO
$O^{(1)}_{\cC}$ decays as an area law like the ordinary confined
phase of a $Z_2$ lattice gauge theory, and there is a SSB of the
subsystem symmetries $Z_{2}^{(\mathrm{sub})}$. The most convenient
way to study this limit, is to take the dual Hamiltonian of
Eq.~\ref{z22d1}, which still has subsystem
$\tZ_2^{(\mathrm{sub})}$ symmetries: \beqn H_d = \sum_{\btx} - K
\tau^1_{\btx} - 2 J \tau^3_{\btx}\tau^3_{\btx + \hx}\tau^3_{\btx +
\hy}\tau^3_{\btx + \hx + \hy}. \label{z22d2} \eeqn The duality
mapping between $\sigma^i$ and $\tau^i$ is the same as the
standard $2d$ Ising-Gauge duality discussed in the previous
section. $\tZ_2^{(\mathrm{sub})}$ inherits and contains
$Z_2^{(\mathrm{sub})}$, but is slightly larger:
$\tZ_2^{(\mathrm{sub})}$ includes another $\tZ_2$ element which
changes the sign of all $\tau^3_{\btx}$. This extra $\tZ_2$
element is the dual of $Z_2^{(1)}$, and it does not change
$\sigma^1_l$ in Eq.~\ref{z22d1}.

The ODO of $\tZ_2^{(\mathrm{sub})}$ is a product of $\tau^3$ on
four corners of a rectangle: \beqn \tO^{(\mathrm{sub})}_{x,y} =
\tau^3_{0,0} \tau^3_{x,0} \tau^3_{0,y} \tau^3_{x,y}.
\label{odo2dsub} \eeqn The ODO defined above is also a product of
the $J$ term in Eq.~\ref{z22d1} within the rectangle. In the
original topological order $K \gg J$, $\tO^{(\mathrm{sub})}_{x,y}$
can be computed through a perturbation of $J/K$, and it decays as
an exponential of the area of the rectangle; while at the SSB
phase of $\tZ_2^{(\mathrm{sub})}$ ($K \ll J$),
$\tO^{(\mathrm{sub})}_{x,y}$ has long range expectation
value~\cite{xumoore}.

Like the previous section, we can embed the dual model
Eq.~\ref{z22d2} into a model with $\tU(1)^{(\mathrm{sub})}$
symmetry: \beqn H_d = \int d^2 \tx \ \frac{U}{2} \hn(\btx)^2 - t
\cos\left(\nabla_x \nabla_y\htheta(\btx)\right) - g \cos\left(
2\htheta(\btx) \right). \label{u12d1} \eeqn The relation between
the operators of the $\tZ_2^{(\mathrm{sub})}$ theory
Eq.~\ref{z22d2} and the $\tU(1)^{(\mathrm{sub})}$ theory
Eq.~\ref{u12d1} is \beqn \tau^x_{\btx} = \exp\left( \ii \pi
\hn(\btx) \right), \ \ \ \ \tau^z_{\btx} = \exp\left( \ii
\htheta(\btx) \right) \eeqn When $g$ is relevant, it will break
the $\tU(1)^{(\mathrm{sub})}$ down to $\tZ_2^{(\mathrm{sub})}$.

However, as was studied before~\cite{paramekanti}, the $g$ term
can only flow strong and become nonperturbative under
renormalization group through ``assistance" from some other terms
such as $\gamma(2 \nabla_\mu \theta)$. If we tune $\gamma$ to
zero, then there exists a stable gapless phase of the model
Eq.~\ref{u12d1} with a larger $\tU(1)^{(\mathrm{sub})}$ symmetry,
and the $g$ term is irrelevant. And in this gapless phase the
system is described by the following action: \beqn \mathcal{S}_d =
\int d\tau d^2 \tx \ \frac{1}{2U} (\partial_\tau \theta)^2 +
\frac{t}{2} (\nabla_x \nabla_y \theta)^2, \label{u12d2} \eeqn
where $\theta$ can be viewed as a free boson instead of a compact
boson. The $\tU(1)^{(\mathrm{sub})}$ reads \beqn \theta(\btx)
\rightarrow \theta(\btx) + f(\tx) + g(\ty). \label{subu1}\eeqn
This gapless phase can also be described by a $U(1)$ gauge theory,
which can be viewed as the parent theory where the original $Z_2$
lattice gauge theory Eq.~\ref{z22d1} is embedded to: \beqn H =
\int d^2x \ \frac{U}{2} (\vec{\nabla} \times \hva)^2 +
\frac{t}{4}\left( (\nabla_x \he_x)^2 + (\nabla_y \he_y)^2 \right).
\label{u12dqed} \eeqn

In this gapless phase, the expectation value of the ODO of the
original $Z_2$ gauge theory $O^{(1)}_\mathcal{C}$ will depend on
the shape of $\mathcal{C}$, but it no longer follows the ``corner
law" Eq.~\ref{corner} of the gapped topological ordered phase $K
\gg J$ in Eq.~\ref{z22d1}. In the gapless phase, the ODO
$O^{(1)}_\cC$ can be written as \beqn \langle O^{(1)}_\mathcal{C}
\rangle = \langle \prod_{\btx \in \cA,
\partial \cA = \cC} \tau^1_{\btx} \rangle \sim \langle
e^{\sum_{\btx \in \mathcal{A}} \ii \pi \hn(\btx)} \rangle. \eeqn

In order to evaluate $\langle O^{(1)}_\mathcal{C} \rangle $ we
will make use of another duality of Eq.~\ref{u12d1} and
Eq.~\ref{u12d2}: \beqn H_{d2} = \int d^2x \ \frac{U}{2} (\nabla_x
\nabla_y \hphi(\bx))^2 - t \cos\left(\hN(\bx)\right). \eeqn Now
$\hphi(\bx)$ and $\hN(\bx)$ are still defined on the sites of the
original lattice $\bx$ (Fig.~\ref{clattice}): \beqn \nabla_x
\nabla_y \htheta(\btx) = - \hN(\bx), \ \ \ \nabla_x \nabla_y
\hphi(\bx) = \hn(\btx). \eeqn The gapless phase has a new dual
description in terms of the continuum limit model of $\hphi(\bx)$:
\beqn \mathcal{S}_{2d} = \int d^2x d\tau \
\frac{1}{2t}(\partial_\tau \phi)^2 + \frac{U}{2} (\nabla_x
\nabla_y\phi)^2. \label{u12d3} \eeqn

In this gapless phase, if we consider a loop $\mathcal{C}$ which
is a rectangle with four corners at $(0,0)$, $(x,0)$, $(0,y)$,
$(x,y)$ (Fig.~\ref{clattice}), the expectation value
$O^{(1)}_\mathcal{C}$ is \beqn && \langle O^{(1)}_\mathcal{C}
\rangle = \langle \prod_{\btx \in \cA,
\partial \cA = \cC} \tau^1_{\btx} \rangle \sim \langle
\exp\left( \sum_{\btx \in \mathcal{A}} \ii \pi \hn(\btx) \right)
\rangle \cr\cr &=& \langle \exp\left( \ii \pi (\hphi_{0,0} -
\hphi_{x,0} - \hphi_{0,y} + \hphi_{x,y}) \right) \rangle \cr\cr
&\sim& \exp\left( \pi^2 (\langle \hphi_{0,0} \hphi_{x,0} \rangle +
\langle \hphi_{0,0} \hphi_{0,y} \rangle + \langle \hphi_{x,y}
\hphi_{x,0} \rangle + \langle \hphi_{x,y} \hphi_{0,y} \rangle -
\langle \hphi_{0,0} \hphi_{x,y} \rangle - \langle \hphi_{0,y}
\hphi_{x,0} \rangle) \right) \cr\cr &\sim& \exp\left( - c \pi^2
\sqrt{\frac{t}{U}} \ \log|x|\log|y| \right).
\label{doublelog}\eeqn This is a faster decay compared with the
corner law in the gapped topologically ordered phase $K \gg J$ in
Eq.~\ref{z22d1}. In the same gapless phase, the expectation value
of $\tO^{\mathrm{sub}}_{x,y}$ defined in Eq.~\ref{odo2dsub} decays
in a similar way as Eq.~\ref{doublelog}, rather than a long range
expectation value as the phase $K \ll J$. Hence this gapless phase
described by Eq.~\ref{u12d2}, Eq.~\ref{u12dqed}, Eq.~\ref{u12d3}
can be viewed as a symmetric phase for both $Z_2^{(1)}$ and
$\tZ^{\mathrm{sub}}_2$ symmetries.

The special double logarithmic scaling in Eq.~\ref{doublelog}
arises from the subsystem symmetries Eq.~\ref{subu1} of the parent
$U(1)$ theory. More technically, in order to evaluate
$O^{(1)}_\mathcal{C}$, we need to compute the equal-time
correlation function $\langle \hphi_{0,0} \hphi_{x,y} \rangle$,
which in the momentum space is~\cite{paramekanti} $G_{k_x, k_y}
\sim \int d\omega \ \omega t/(\omega^2 + t U k_x^2 k_y^2 ) \sim
1/|k_xk_y|$. The double linear divergence at $k_x \rightarrow 0$
and $k_y \rightarrow 0$ leads to the special double logarithmic
scaling in real space. The results of this subsection is
summarized in the table below.

\begin{center}
\begin{tabular}{ |c|c|c|c|}
 \hline  Special $2d$ $Z_2$ Gauge theory Eq.~\ref{z22d1}  & $K \gg J$ & $K \ll J$ & Gapless Phase
 \\
 \hline
 $O^{(1)}_{\cC}$  & Corner law & Area law & $\exp\left( - c \pi^2 \sqrt{t/U} \ \log|x|\log|y|
 \right)$ for rect. $\cC$
 \\
 \hline
 $\tO^{\mathrm{sub}}_{x,y}$ & Area law & Long range & $\exp\left( - \tilde{c} \pi^2 \sqrt{U/t} \ \log|x|\log|y| \right)$
 \\
 \hline

\end{tabular}
\label{2dsub}
\end{center}

\subsection{3d Example}
\label{sub3d}

\begin{figure}
\includegraphics[width=0.35\textwidth]{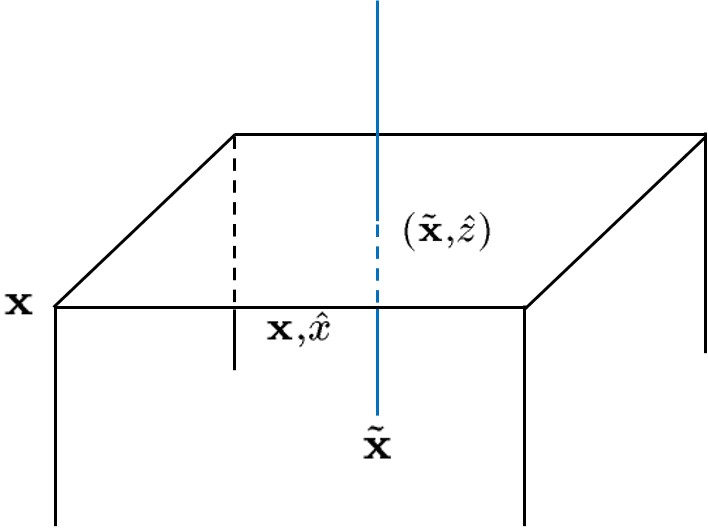}
\caption{The cubic lattice and the dual lattice for models
considered in section~\ref{sub3d}.} \label{clattice3d}
\end{figure}

We now consider a $3d$ $Z_2$ lattice gauge theory defined on the
cubic lattice, which has both the 1-form symmetry, and subsystem
symmetries: \beqn H &=& \sum_{\bx, \hmu, \hnu} - K \sigma^3_{\bx,
\hmu} \sigma^3_{\bx, \hnu} \sigma^3_{\bx + \hmu, \hnu}
\sigma^3_{\bx + \hnu, \hmu} - J \sigma^1_{\bx, \hmu} \sigma^1_{\bx
+ \hmu, \hmu} \cr\cr \leftrightarrow \ \ H_d &=& \sum_{\btx, \hmu,
\hnu} - K \tau^1_{\btx, \hmu} - \sum_{\hrho \perp \hmu,\hnu} J
\hB_{\btx, \hmu\hnu} \hB_{\btx + \hrho,\hmu\hnu}. \label{z23d2}
\eeqn where $\hB_{\btx, \hmu\hnu} = \tau^3_{\btx, \hmu}
\tau^3_{\btx, \hnu} \tau^3_{\btx + \hmu, \hnu} \tau^3_{\btx +
\hnu, \hmu}$. The theory $H$ has an ordinary $Z_2^{(1)}$ symmetry
like Eq.~\ref{z23d1}, plus subsystem symmetries with conserved
quantities: \beqn \Sigma_{\hx; (y,z)} = \prod_{y,z =
\mathrm{Const}} \sigma^3_{\bx, \hx}, \ \ \ \ \Sigma_{\hy; (x, z)}
= \prod_{x,z = \mathrm{Const}} \sigma^3_{\bx, \hy}, \ \ \ \
\Sigma_{\hz; (x,y)} = \prod_{x,y = \mathrm{Const}} \sigma^3_{\bx,
\hz}. \label{subcon1} \eeqn $x, y, z$ are the three coordinates of
$\bx$. The ODO of the $Z_2^{(1)}$ 1-form symmetry is the same as
Eq.~\ref{z23d1}: $O^{(1)}_\cC = \prod_{l \in \cC} \sigma^3_{l}$.
Due to the extra subsystem conserved quantities in
Eq.~\ref{subcon1}, and since $O^{(1)}_{\cC}$ is a product of
segments of these extra conserved quantities, the expectation
value of $O^{(1)}_{\cC}$ in the phase $K \gg J$ also decays with a
corner law, i.e. the expectation value of $O^{(1)}_{\cC}$ decays
only when $\cC$ takes a turn; in the phase $K \ll J$, there is a
SSB of the subsystem symmetry, and the expectation value of
$O^{(1)}_{\cC}$ decays with an area law.

The dual Hamiltonian $H_d$ has the same $\tZ_2^{(1)}$ symmetry as
the dual of the ordinary $Z_2$ quantum lattice gauge theory, with
extra subsystem symmetries as well. The ODO we will consider for
$H_d$ is \beqn \tO^{(1)}_{\cC, \cC'} = \prod_{\tl \in \cC}
\tau^3_{\tl} \prod_{\tl \in \cC'} \tau^3_{\tl}.
\label{odo3d2}\eeqn There are still subsystem symmetries in $H_d$
of Eq.~\ref{z23d2}, with conserved subsystem symmetry charges such
as \beqn \tilde{\Sigma}_{\hz; (\ty,\tz)} = \prod_{\ty,\tz =
\mathrm{Const}} \tau^1_{\btx, \hz}, \ \ \ \ \tilde{\Sigma}_{\hz;
(\tx,\tz)} = \prod_{\tx, \tz = \mathrm{Const}} \tau^1_{\btx, \hz},
\ \ \ \cdots \eeqn These conserved subsystem charges are not
entirely independent from each other due to the Gauss-law gauge
constraint imposed on $\tau^1$. Due to these subsystem symmetries
in the dual model, we restrict our discussions to the cases when
$\cC$ and $\cC'$ in $\tO^{(1)}_{\cC, \cC'}$ are completely
parallel with each other, and separated along the direction
orthogonal to both loops, (for example, $\cC$ and $\cC'$ can be
identical squares in two XY planes, but separated along the $\hz$
direction), because only then would the ODO commute with all the
conserved 1-form charges of the dual model Eq.~\ref{z23d1}, and
also commute with the subsystem conserved charges
$\tilde{\Sigma}$. When $\cC$ and $\cC'$ are identical loops in XY
planes separated along the $\hz$ direction, $\tO^{(1)}_{\cC,
\cC'}$ is also a product of $J\sigma^1_{\bx, \hz}\sigma^1_{\bx +
\hz, \hz}$ in $H$ of Eq.~\ref{z23d2} within the $3d$ region
sandwiched between $\cC$ and $\cC'$; while $O^{(1)}_\cC$ is still
a product of the $K$ term enclosed by $\cC$.
%This expectation value
%of $\tO^{(1)}_{\cC, \cC'}$ is nonzero when $\cC$ and $\cC'$ are
%both in the XY planes, but separated parallelly along the $z$
%axis.

In the phase $K \ll J$, the expectation value of $\tO^{(1)}_{\cC,
\cC'}$ can again be computed through a perturbation of $K/J$: it
decays as a perimeter law of $\cC$ (or equivalently $\cC'$), but
it does not decay with the distance between $\cC$ and $\cC'$. In
the phase $K \gg J$, the expectation value of $\tO^{(1)}_{\cC,
\cC'}$ would decay exponentially with the distance between $\cC$
and $\cC'$, and also exponentially with the area of $\cC$ (or
$\cC'$).

It is unknown whether model Eq.~\ref{z23d2} has a second order
transition between the two phases mentioned above or not. But
again we can embed the models into a parent model with
$U(1)^{(1)}$ and $\tU(1)^{(1)}$ symmetries. For instance, the
$H_d$ in Eq.~\ref{z23d2} can be embedded into \beqn H_d = \int d^3
\tx \ \sum_\mu \frac{U}{2} \he_{\btx, \hmu}^2 - t \cos \left(
\nabla_z (\nabla_x \ha_y - \nabla_y \ha_x) \right) +
(\mathrm{permute} \ x,y,z) - g \cos(2 \ha_\mu). \label{u13d1}
\eeqn $\he$ and $\ha$ are defined on the dual lattice links
$(\btx, \hmu)$, which are also the plaquettes of the original
cubic lattice (Fig.~\ref{clattice3d}). This model admits a gapless
phase described by the following action: \beqn \mathcal{S}_d =
\int d^3 \tx d\tau \ \frac{1}{2U} (\partial_\tau \va)^2 +
\frac{t}{2} \left( \nabla_z (\nabla_x a_y - \nabla_y a_x)
\right)^2 + (\mathrm{permute} \ x,y,z). \label{u13d2}  \eeqn In
this gapless phase, the ODO Eq.~\ref{odo3d2} becomes \beqn
\tO^{(1)}_{\cC, \cC'} = \prod_{\tl \in \cC} \tau^3_{\tl}
\prod_{\tl \in \cC'} \tau^3_{\tl} \sim \exp\left( \ii \oint_{\cC}
\ha_{\mu} dx^\mu \right) \exp\left( - \ii \oint_{\cC'} \ha_{\nu}
dx^\nu \right). \eeqn The expectation value of $\tO^{(1)}_{\cC,
\cC'}$ can be evaluated with the continuum limit action
Eq.~\ref{u13d2}.

Our goal is to show that, the behavior of $\tO^{(1)}_{\cC,\cC'}$
is different from the gapped phases. This effect can be readily
shown by considering the simple case when both $\cC$ and $\cC'$
are unit plaquettes in the XY planes, separated in the $z$
direction by distance $Z$. Then \beqn \tO^{(1)}_{\cC, \cC'}(Z)
&\sim& \exp \left( \langle (\nabla_x \ha_{y} - \nabla_y
\ha_{x})_{z = 0} (\nabla_x \ha_y - \nabla_y \ha_x)_{z = Z} \rangle
\right) \cr\cr &\sim& \exp\left(- c_1 \sqrt{\frac{U}{t}} \log Z
\right) \sim \frac{1}{|Z|^{2\Delta_{\cC,\cC'}}}, \ \ \
\Delta_{\cC,\cC'} \sim \sqrt{\frac{U}{t}}. \eeqn This power-law
decay along the $z$ direction originates from the fact that the
correlation function $\langle (\nabla_x \ha_{y} - \nabla_y
\ha_{x})_{z = 0} (\nabla_x \ha_y - \nabla_y \ha_x)_{z = Z}
\rangle$ has a singularity $1/k_z$ in the momentum space near $k_z
= 0$. This power-law scaling along $z$ is already very different
from the expectation value of $\tO^{(1)}_{\cC, \cC'}(Z)$ in the
gapped phases of the models in Eq.~\ref{z23d2}. We also made
efforts to compute $\tO^{(1)}_{\cC,\cC'}$ for $\cC$, $\cC'$ with
more general shapes, this calculation is presented in the
appendix.

To evaluate $O^{(1)}_\cC$, again it is more convenient to use a
third dual description of the theory: \beqn H_{d2} = \int d^3 x \
\frac{U}{2} \left(\nabla_x \nabla_y (\hphi_x(\bx) - \hphi_y(\bx))
\right)^2 - t \cos\left(\hN_z(\bx)\right) + (\mathrm{permute} \
x,y,z) \label{u13d3} \eeqn The operators in Eq.~\ref{u13d3} are
related to the operators in Eq.~\ref{u13d1} through the mapping
(the duality between $H_d$ and $H_{d2}$ was first discussed in
Ref.~\onlinecite{xufisher}) \beqn && \he_{\btx,\hz} = \nabla_x
\nabla_y (\hphi_x (\bx) - \hphi_y (\bx)), \ \ \mathrm{and \
permutation \ of} \ x, y, z. \cr \cr && \nabla_z (\nabla_x
\ha_{\btx, \hy} - \nabla_y \ha_{\btx, \hx}) = - \hN_z(\bx), \ \
\mathrm{and \ permutation \ of} \ x, y, z. \eeqn The gapless phase
is described by the following action: \beqn \mathcal{S}_{d2} =
\int d^3 x d\tau \ \frac{U}{2} (\nabla_x \nabla_y (\phi_x -
\phi_y))^2 + \frac{1}{2t} (\partial_\tau \phi_z)^2 +
(\mathrm{permute} \ x,y,z) \label{u13d4} \eeqn $\hphi_{i}(\bx)$
and $\hN_i(\bx)$ are three pairs of conjugate variables defined on
the sites of the original cubic lattice $\bx$. Let us assume that
the loop $\cC$ in $O^{(1)}_\cC$ is a rectangle in the XY plane,
then \beqn O^{(1)}_\cC &=& \prod_{l \in \cC} \sigma^3_l =
\prod_{(\btx, \hz) \in \cA} \tau^1_{\btx, \hz} = \prod_{(\btx,
\hz) \in \cA} \exp(\ii \pi \he_{\btx, \hz}) \cr\cr &=& \prod_{\bx
\in \cA} \exp\left(\ii \pi \nabla_x \nabla_y (\hphi_x (\bx) -
\hphi_y (\bx)) \right) = \exp\left(\ii \pi \sum_{\bx \in \cA}
\nabla_x \nabla_y (\hphi_x (\bx) - \hphi_y (\bx)) \right) \cr\cr
&=& \exp\left(\ii \pi (\hphi_x(0,0) - \hphi_x(x,0) - \hphi_x(0,y)
+ \hphi_x(x,y)) - \ii \pi (\hphi_y(0,0) - \hphi_y(x,0) -
\hphi_y(0,y) + \hphi_y(x,y))\right). \eeqn Again since our goal is
to show that $O^{(1)}_\cC$ has different behavior from the two
gapped phases $K \gg J$ and $K \ll J$, it is sufficient to
consider a special ``narrow rectangular" shape of $\cC$, i.e. $y =
1$, but $x \gg 1$. $\langle O^{(1)}_\cC \rangle $ in this case is
evaluated as $\exp(\pi^2 \langle \nabla_y(\phi_x -
\phi_y)_{0,0}\nabla_y(\phi_x - \phi_y)_{x,0} \rangle) $. The key
correlation function $\langle \nabla_y(\phi_x -
\phi_y)_{0,0}\nabla_y(\phi_x - \phi_y)_{x,0} \rangle$ has an
infrared singularity as $1/|k_x|$ near $k_x = 0$. $O^{(1)}_\cC$
with a narrow rectangular $\cC$ scales as \beqn \langle
O^{(1)}_\cC \rangle \sim \frac{1}{|x|^{\Delta_{\cC}}}, \ \ \ \ \
\Delta_{\cC} \sim \sqrt{\frac{t}{U}}. \eeqn The power law decay of
$O^{(1)}_\cC$ is very different from the two gapped phases of
Eq.~\ref{z23d2}. The results of this subsection are summarized in
the table below.

\begin{center}
\begin{tabular}{ |c|c|c|c|}
 \hline  Special $3d$ $Z_2$ Gauge theory Eq.~\ref{z23d2}  & $K \gg J$ & $K \ll J$ & Gapless Phase
 \\
 \hline
 $O^{(1)}_{\cC}$ with rect. $\cC$ in XY & Corner law & Area law & $\frac{1}{|x|^{\Delta_{\cC}}}$, with $y = 1$ and $x \gg 1$.
 \\
 \hline
 $\tO^{(1)}_{\cC, \cC'}$ parallel $\cC$, $\cC'$ in XY; & Area law of $\cC$, $\cC'$; & Perimeter law of $\cC$; & $\frac{1}{|Z|^{2\Delta_{\cC,\cC'}}}$, for unit square
 \\
 separated along $\hz$& exponential decay with $Z$ & long range with $Z$ &  $\cC,\cC'$ separated along $z$
 \\
 \hline
\end{tabular}
\label{3dsub}
\end{center}

\section{Summary}

In this manuscript we analyzed the behavior of order diagnosis
operators (ODO), at fine-tuned critical points or gapless phases
of lattice systems with microscopic discrete categorical
symmetries. The symmetries on both sides of the duality of the
lattice models are constituents of the entire categorical symmetry
of the system. We demonstrate that at these selected
criticalities, the behavior of ODOs of categorical symmetries can
be evaluated analytically, and they could have rather different
scalings from the gapped phases of these models, where the ODO can
be computed using the perturbation theory. The existence of
subsystem symmetries of some of the systems intrinsically modify
the behavior of ODOs at both the gapped phases, and the
criticalities. And in examples with subsystem symmetries, we found
that at these criticalities the scaling of ODOs defined on both
sides of the duality of the lattice models is substantially
different from the gapped phases.

While preparing for our manuscript, we became aware of a work that
numerically computed the behavior of ODO of $\tZ_2^{(1)}$ at the
$3D$ Ising critical point, combined with theoretical comparison
with free field theories~\cite{chengcat}. The conclusion in this
work is that, the $\tZ_2^{(1)}$ symmetry is still spontaneously
broken at the $3D$ Ising critical point. The conclusion is similar
to ours at the fine-tuned Lifshitz criticality of $2d$ lattice
quantum Ising systems.

This work is supported by NSF Grant No. DMR-1920434, the David and
Lucile Packard Foundation, and the Simons Foundation. The authors
thank Chao-Ming Jian for helpful discussions.

\appendix

\section{More calculations for $\tO^{(1)}_{\cC,\cC'}$}

Let's first analyze the expectation value of $\tO^{(1)}_{\cC,
\cC'}$ defined in Eq.~\ref{odo3d2}, which can be calculated using
the continuous gauge theory Eq.~\ref{u13d2} via \beqn
\left\langle \!\!\right.\widetilde{O}_{\mathcal{C},\mathcal{C}^{\prime}}^{(1)}\left.\!\!\right\rangle \sim\exp\left[\left(\oint_{\mathcal{C}}\oint_{\mathcal{C}^{\prime}}-\frac{1}{2}\oint_{\mathcal{C}}\oint_{\mathcal{C}}-\frac{1}{2}\oint_{\mathcal{C}^{\prime}}\oint_{\mathcal{C}^{\prime}}\right)\left\langle a_{\mu}(\mathbf{x})a_{\nu}(\mathbf{x}^{\prime})\right\rangle dx^{\mu}dx^{\prime\nu}\right].
\eeqn With a Faddeev-Popov type
gauge fixing $\zeta$ term, the gauge field propagator
$D_{\mu\nu}(\omega,\mathbf{k})$ is given by \beqn
D_{\mu\nu}^{-1}(\omega,\mathbf{k})&=\left(\begin{array}{ccc}
\omega^{2}/U+2tk_{y}^{2}k_{z}^{2} & -tk_{x}k_{y}k_{z}^{2} & -tk_{x}k_{z}k_{y}^{2}\\
-tk_{x}k_{y}k_{z}^{2} & \omega^{2}/U+2tk_{z}^{2}k_{x}^{2} & -tk_{y}k_{z}k_{x}^{2}\\
-tk_{x}k_{z}k_{y}^{2} & -tk_{y}k_{z}k_{x}^{2} &
\omega^{2}/U+2tk_{x}^{2}k_{y}^{2}
\end{array}\right)-\zeta^{-1}k_{\mu}k_{\nu}.
\eeqn Our gauge choice is $\zeta\rightarrow0$. Since we are
interested in the expectation value of ODO of a static state, we
will use the equal time Green's function. Directly using the full
form of $D_{\mu\nu}$ would be tedious, but we observe that
$D_{xx}$ has linear singularity at $k_y \rightarrow 0$, and $k_z
\rightarrow 0$, which will dominate IR behavior of the Green's
function. We can extract the most singular part of the Green's
function, then $D_{xx}$ at $\tau = 0$ reads \beqn
D_{xx}(\tau=0,\mathbf{k})=\int\frac{d\omega}{2\pi}
D_{xx}(\omega,\mathbf{k})= \sqrt{\frac{U}{t}}\frac{1}{\sqrt{6}}
\left(\frac{k_{y}^{2}}{\left(k_{x}^{2}+k_{y}^{2}\right){}^{3/2}}\frac{1}{\left|k_{z}\right|}
+\frac{k_{z}^{2}}{\left(k_{x}^{2}+k_{z}^{2}\right){}^{3/2}}\frac{1}{\left|k_{y}\right|}\right)
+ \cdots \label{greensing} \eeqn This approximate form of Green's
function captures the singularity at $k_y \rightarrow 0$ and $k_z
\rightarrow 0$ separately. There is an extra singularity when
multiple momenta approach zero simultaneously. But since this
extra singularity occurs at a much smaller measure of the momentum
space compared with the singularities captured by
Eq.~\ref{greensing}, we take the approximate form of Green's
function Eq.~\ref{greensing}. Further analysis may be demanded to
address all singularities in the Green's function.

Similarly, we approximate the off-diagonal term $D_{xy}$ around
its singularity $k_{z}=0$ \beqn
D_{xy}(\tau=0,\mathbf{k})=\sqrt{\frac{U}{t}}
\frac{-k_{x}k_{y}}{\sqrt{6}\left(k_{x}^{2}+k_{y}^{2}\right){}^{3/2}}\frac{1}{\left|k_{z}\right|}
+ \cdots. \eeqn Other components of $D_{\mu\nu}$ can be obtained
by the permutations of $k_x, k_y, k_z$. The real-space expression
of the Green's function is then obtained through Fourier
transformation:
\begin{equation}
D_{\mu\nu}(\tau=0,\mathbf{x})=\sqrt{\frac{U}{t}}\frac{-1}{2\sqrt{6}\pi^{2}}\left(\begin{array}{ccc}
\frac{x^{2}\log\left|z\delta\right|}{(x^{2}+y^{2})^{3/2}}+\frac{x^{2}\log\left|y\delta\right|}{(x^{2}+z^{2})^{3/2}} & \frac{xy\log\left|z\delta\right|}{(x^{2}+y^{2})^{3/2}} & \frac{xz\log\left|y\delta\right|}{(x^{2}+z^{2})^{3/2}}\\
\frac{xy\log\left|z\delta\right|}{(x^{2}+y^{2})^{3/2}} & \frac{y^{2}\log\left|x\delta\right|}{(y^{2}+z^{2})^{3/2}}+\frac{y^{2}\log\left|z\delta\right|}{(y^{2}+x^{2})^{3/2}} & \frac{yz\log\left|x\delta\right|}{(y^{2}+z^{2})^{3/2}}\\
\frac{xz\log\left|y\delta\right|}{(x^{2}+z^{2})^{3/2}} &
\frac{yz\log\left|x\delta\right|}{(y^{2}+z^{2})^{3/2}} &
\frac{z^{2}\log\left|y\delta\right|}{(z^{2}+x^{2})^{3/2}}+\frac{z^{2}\log\left|x\delta\right|}{(z^{2}+y^{2})^{3/2}}
\label{Green of a}
\end{array}\right),
\end{equation}
where $\delta>0$ is a small IR cut-off, which is needed in the
Fourier transformation of $1/\left|k\right|$. This is the
effective real-space Green function that can be used to calculate
the scaling behaviors of
$\langle\tilde{O}_{\mathcal{C},\mathcal{C}^{\prime}}^{(1)}\rangle$.

Let's consider two identical squares
$\mathcal{C},\mathcal{C}^{\prime}$ that are completely parallel to
each other. We assume $\mathcal{C}$ has four corners
$\left(0,0,0\right),\left(L,0,0\right),\left(L,L,0\right),\left(0,L,0\right)$,
and $\mathcal{C}^{\prime}$ has four corners
$\left(0,0,Z\right),\left(L,0,Z\right),\left(L,L,Z\right),\left(0,L,Z\right)$.
Based on the real-space Green function Eq. \ref{Green of a}, an
integral over $\mathcal{C},\mathcal{C}^{\prime}$ leads to
\begin{equation}
-\log\langle\tilde{O}_{\mathcal{C},\mathcal{C}^{\prime}}^{(1)}\rangle=\sqrt{\frac{U}{t}}\frac{4L}{\sqrt{6}\pi^{2}}\left(\begin{array}{c}
\left(2(\sqrt{L^{2}+Z^{2}}-Z)/L+\log(\sqrt{L^{2}+Z^{2}}-L)\right)\log(L/\epsilon)+\log L\left(\log L-3\log\epsilon\right)\\
-\log(LZ)+\left(\sqrt{2}-\sinh^{-1}(1)\right)\log(Z/\epsilon)+2\log\epsilon(\log\epsilon+1)
\end{array}\right).
\end{equation}
where $\epsilon>0$ is a small UV cut-off. It is important to
notice that, although the real space Green's function has a
dependence on the IR cut-off $\delta$, the final result of
$\tO^{(1)}_{\cC,\cC'}$ is free from any IR-divergence. We are most
interested in the behaviors of
$\langle\tilde{O}_{\mathcal{C},\mathcal{C}^{\prime}}^{(1)}\rangle$
under the large-$L$ and large-$Z$ limits:
\begin{equation}
\langle\tilde{O}_{\mathcal{C},\mathcal{C}^{\prime}}^{(1)}\rangle\sim\begin{cases}
e^{-\sqrt{\frac{U}{t}}\frac{4}{\sqrt{6}\pi^{2}}L\left(\log(L/\epsilon)+\sqrt{2}-1-\sinh^{-1}(1)\right)\log Z}=e^{-c_{1}\sqrt{\frac{U}{t}}\log Z} & L<+\infty,Z\rightarrow+\infty\\
e^{-\sqrt{\frac{U}{t}}\frac{4}{\sqrt{6}\pi^{2}}\left(2\log(Z/\epsilon)+1-\log2\right)L\log L}=e^{-c_{2}\sqrt{\frac{U}{t}}L\log L} & Z<+\infty,L\rightarrow+\infty
\end{cases},
\end{equation}
where $c_{1}$ and $c_{2}$ are two numerical coefficients which
depend on the UV cut-off $\epsilon$.

\bibliography{cnote}

\end{document}